\documentclass{eas}
\usepackage{graphicx}
\newcommand{\Ms}{$M_{\odot}$}
\newcommand{\rad}{$R_{\odot}$}
\newcommand{\rstar}{$R_{\star}$}
\begin{document}

%%-----------------------------
%%      the top matter
%%-----------------------------
\title{The formation of Polycyclic Aromatic Hydrocarbons in evolved circumstellar environments } 
\runningtitle{PAHs in evolved circumstellar environments}

\author{Isabelle Cherchneff}\address{Departement Physik, Universit{\"a}t Basel, Klinglebergstrasse 82, CH-4056 Basel, Switzerland}
%\author{...}\address{...}
%\author{...}\address{...}
%
%
\begin{abstract}

The formation of Polycyclic Aromatic Hydrocarbons in the circumstellar outflows of evolved stars is reviewed, with an emphasis on carbon stars on the Asymptotic Giant Branch. Evidence for PAHs present in their winds is provided by meteoritic studies and recent observations of the Unidentified Infrared bands. We detail the chemical processes leading to the closure of the first aromatic ring as well as the growth mechanisms leading to amorphous carbon grains. Existing studies on PAH formation in evolved stellar envelopes are reviewed and new results for the modelling of the inner wind of the archetype carbon star IRC+10216 are presented. Benzene, C$_6$H$_6$, forms close to the star, as well as water, H$_2$O, as a result of non-equilibrium chemistry induced by the periodic passage of shocks. The growth process of aromatic rings may thus resemble that active in sooting flames due to the presence of radicals like hydroxyl, OH. Finally, we discuss possible formation processes for PAHs and aromatic compounds in the hydrogen-rich R CrB star, V854 Cen, and their implication for the carriers of the Red Emission and the Diffuse Interstellar Bands.  
\end{abstract}
\maketitle
%%-----------------------------
%%      your text
%%-----------------------------
\section{Introduction}
Upon the suggestion by L{\'e}ger \& Puget (1984) and Allamandola, Tielens \& Barker (1985) that polycyclic aromatic hydrocarbons (PAHs) were responsible for the emission of a series of unidentified infrared (UIR)  bands ubiquitous to the interstellar medium (ISM), large efforts have been devoted to understand the presence of these aromatic molecules in space. PAHs are commonly found on Earth in the exhaust fumes of car engines and are key intermediates in the inception and growth of soot particles in incomplete combustion processes (Haynes \& Wagner 1981). They  are associated to the fine particles called aerosols released by human activities in the Earth troposphere. PAHs are also mutagens and thus carcinogens. Therefore, PAHs are extensively studied on Earth, providing an abundant and exhaustive scientific literature of great interest to the study of PAHs in space. Cosmic PAHs are observed through their UIR bands in several local, low- and high-redshift environments including HII regions, Proto- and Planetary Nebulae, young stellar objetcs, and the ISM of high redshift galaxies (Tielens 2008). We can define two populations of PAHs in space: 1) PAHs formed directly from the gas phase in dense media, and 2) PAHs as products of dust sputtering in harsh environments. These two distinct classes require very diverse environments and processes to form, the second class being prevalent in galaxies and most readily observed through the detection of the UIR bands. In this review, we focus on the first class of PAHs, those forming under specific conditions from the gas phase in evolved circumstellar environments. We discuss evidence for PAHs in stardust inclusions in meteorites and PAHs observed in evolved low-mass stars. We detail the formation routes to benzene and its growth to several aromatic rings. We review existing models and present new results on benzene formation in the extreme carbon star IRC+10216. Finally, the PAH synthesis in R CrB stars is discussed. 

\section{Carbon dust providers in our galaxy.}

PAHs are intimately linked to the formation of amorphous carbon (AC) dust or soot from the gas phase in combustion processes on Earth. It is thus natural reckoning that similar formation conditions may exist in space that are conducive to the synthesis of PAHs. Those conditions characteristic of the gas phase are: (1) an initial carbon-, hydrogen-rich chemical composition; (2) high densities; (3) and high temperatures. Such conditions are usually found in space in circumstellar environments such as the winds or ejecta of evolved stars. Furthermore, these evolved objects are the prominent providers of dust in our galaxy. Therefore, the formation of PAHs in space from the gas phase is linked to evolved stellar objects synthesising AC dust. The prevalent AC grain makers in our galaxy include the late stages of evolution of low- and high-mass stars. They are the wind of low-mass carbon-rich stars on the Asymptotic Giant Branch (AGB), the ejecta of Type II supernovae, the colliding winds of carbon-rich Wolf-Rayet stellar binaries, and the clumpy winds of R CrB stars. The physical and chemical conditions pertaining to these environments are summarized in Table 1. For carbon AGB stars, the gas layers close to the photosphere that are periodically shocked satisfy the required conditions for PAH formation (Cherchneff 1998). Conversely, the ejecta of supernovae do not as the hydrogen present in the gas is not microscopically mixed within the ejecta layers but is in the form of macroscopically-mixed blobs of gas (Kifonidis et al. 2006). Carbon-rich Wolf-Rayet stars have lost their hydrogen envelopes and their wind is H-free. The observed AC dust is  thus synthesised by a chemical route that does not involve PAHs but rather bare carbon chains and rings (Cherchneff et al. 2000). However, WC stars are part of binary systems in which the companion is usually a OB star characterised by a H-rich wind, thus implying that hydrogen is present in the colliding wind region. But owing to the high gas temperatures, it is unlikely that the chemical routes responsible for PAHs synthesis may be active in these hot regions. Finally, R CrB stars are extremely rare, evolved low-mas stars exhibiting supergiant photospheric conditions and episodically ejecting clumps of gas rich in AC dust. Their wind is H-free but there exists one object, V854 Cen, which is H-rich and for which the UIR bands have been observed with ISO (Lambert et al. 2001). PAHs may therefore play a role in the formation process of AC dust in this specific object. 

\begin{table}
\caption{Most important amorphous carbon (AC) dust stellar providers in our galaxy. Listed in the first column are: (1) the amount of AC dust formed over the star lifetime, (2) the region where dust forms, (3) the presence of hydrogen, (4) the gas density and temperature at which dust forms, and (5) the observed molecules related to AC dust synthesis. Value for supernovae are for SN 1987A (Ercolano et al. 2007) and for the SN progenitor of Cas A (Nozawa et al. 2010). Value for R CrB stars is an upper limit assuming all carbon condenses in AC dust.}            
\label{table:1}      
\centering                          
\begin{tabular}{lc c }  
\\  
\hline\hline       
                      % To combine 4 columns into a single one 
%HJD & $E$ & Method\#2 & \multicolumn{4}{c}{Method\#3}\\ 
Star & Carbon AGB& Type II Supernova \\
\hline                    
AC mass (\Ms)& 3$\times 10^{-3}$ - 1$\times 10^{-2}$& 7$\times 10^{-4}$ - 7$\times 10^{-2}$ \\ 
Dust locus & Shocked inner wind& Ejecta \\
Hydrogen & Yes & Yes - not mixed with C\\
Gas density (cm$^{-3}$)& 10$^8$-10$^{13}$ & 10$^9$-10$^{12}$ \\
Gas temperature (K)& Low: 1000 - 1500 &High: 3000\\
Key species & C$_2$H$_2$ & CO \& SiO\\
\hline\hline
Star &Carbon-rich Wolf-Rayet& R CrB\\ 
\hline 
AC mass (\Ms)& 0.1 & 2$\times 10^{-6}$\\
Dust locus& Colliding winds & Episodic clumps \\
Hydrogen & No & No except V854 Cen \\
Gas density (cm$^{-3}$)& 10$^{10}$ & 10$^9$-10$^{11}$ \\
Gas temperature (K)& High: 3000 & Medium: 1500 - 2500 \\
Key species& No observation yet & C$_2$ \\

\hline                              %inserts single line
\end{tabular}     
\end{table}

Studies on the formation of PAHs in circumstellar environments have focused on the prevalent AC dust makers, the carbon AGB stars, represented by the well-studied carbon star IRC+10216. We will thus concentrate on these stellar dust providers in the rest of this review.   

\section{Evidence for circumstellar PAHs in meteorites}

Crucial information on the chemical composition, structure, and formation locus of stardust has been gained since the isolation of presolar grains in primitive meteorites and their study by laboratory microanalyses. Graphite grains bearing the isotopic composition resulting from the nucleosynthesis at play in low-mass stars have been isolated implying a formation locus in AGB outflows (Zinner 1997). Two basic graphite spherule morphologies are found, designated by 'onion' and cauliflower' (Bernatowicz et al. 2006). Two thirds of the onion-type spherules are characterised by concentric outer shells of well-crystallised graphite surrounding a nanocrystalline carbon core consisting of graphene curled sheets indicative of pentagonal ring insertion. One quarter of this core mass is in the form of small PAHs. Small aromatics including phenantrene (C$_{14}$H$_{10}$) or chrysene (C$_{18}$H$_{12}$)   have further been identified in the study of presolar graphite onion spherules. Those aromatics bore similar isotopic anomalies than their parent graphite circumstellar grains, indicating a possible formation in AGB winds (Messenger et al. 1998). The cauliflower spherules are formed of contorted graphene sheets with no nanocrytalline core. The morphology of these two types of spherules hint at dust condensation scenarios in AGB winds. The onion morphology indicates that the inner gas layers where dust forms are hotter, with long enough residence times to form dust precursors like PAHs and graphene sheets and allow the outer layers of the newly-formed dust grain to become well graphitised. The cauliflower structure on the other hand is indicative of a lower temperature inner wind and rapid motion outward of the shocked layers above the photosphere. Dust precurors like PAHs have formed and are included into graphene sheets but not enough time is given for these amorphous structures to become crystalised. An amorphous, less structured carbon spherule thus results from these milder inner wind conditions. 

Therefore, there exists direct evidence of the presence of PAHs in the dust synthesis zone 
of carbon-rich AGB stars from the study of presolar grains extracted from meteorites. Moreover, it appears that PAHs do form as precursors before the end of the dust condensation process as found in the pyrolysis of hydrocarbons and in combustion processes like hydrocarbon-fueled flames (Frenklach \& Warnatz 1987, Richter \& Howard 2000).  

\section{Evidence for circumstellar PAHs from observations}

The common fingerprints of PAHs in space are the emission of the UIR bands through excitation by a single ultraviolet (UV) photon. These bands can be detected in any circumstellar medium, providing the existence of a UV radiation field. The photospheric temperatures of carbon AGB stars are too low to produce a strong UV stellar radiation field capable of exciting PAH molecules in the dust formation zone. Buss et al. (1991) and Speck \& Barlow (1997) observed the carbon star TU Tauri, part of a binary system with a blue companion providing for UV photons, and detected possible UIR bands. More recently, Boersma et al. (2006) analysed the SWS ISO spectra of several carbon stars including TU Tau and found UIR emission bands in the latter only. They ascribed the UIR spectrum to PAH excitation by the UV radiation field of the companion star when other single carbon stars in the sample were devoid of UIR bands in their spectrum. They pointed out that PAHs may be present in any carbon star but not observable due to the lack of exciting radiation. Another probe to the presence of PAHs in AGB stars is the study of S stars, i.e., oxygen-rich AGB stars on the verge of dredging up carbon in their photosphere to become carbon stars. They are characterised by a photospheric C/O ratio less than but $\simeq$ 1 and by the fact that they do not yet produce carbon dust. Smolders et al. (2010) analysed IR low resolution spectra taken by Spitzer of several galactic S stars. Four of them characterised by C/O ratios very close to 1 show the UIR bands in their spectra. Among those, three stars show weak extended 6.2 $\mu$m emission bands characteristic of PAHs with a mixture of aliphatic and aromatic bonds excited by a weak UV field. S stars possess a dual chemistry in their dust formation zone due to the pulsation-induced shocks as shown by Cherchneff (2006). As a result of non-equilibrium chemistry, O-bearing molecules form close to the stellar photosphere whereas hydrocarbons (e.g. C$_2$H$_2$) form at larger radii ($\sim$ 3 \rad). PAHs may thus form at lower temperatures and densities, favouring small disordered aromatic structures linked by aliphatic bonds. At these radii, the formation of AC dust may be hampered by the low densities and residence time resulting in a population of small aromatic/aliphatic structures that can not turn into dust but are responsible for the observed UIR bands. 

\section{Chemical routes to PAH formation in flames}

\begin{figure} 
\includegraphics[width=6.4cm]{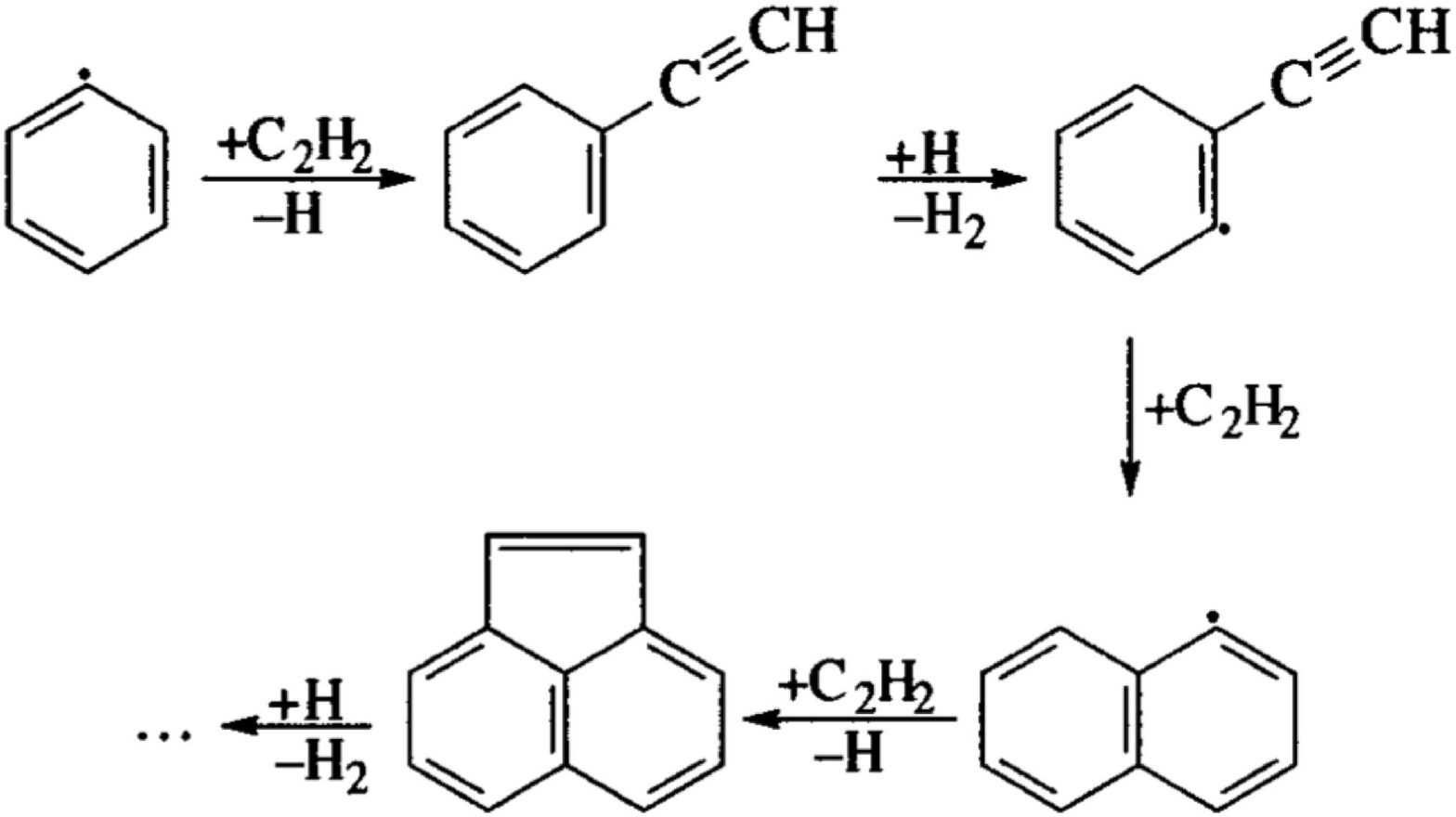}  
\qquad
\includegraphics[width=5.4cm]{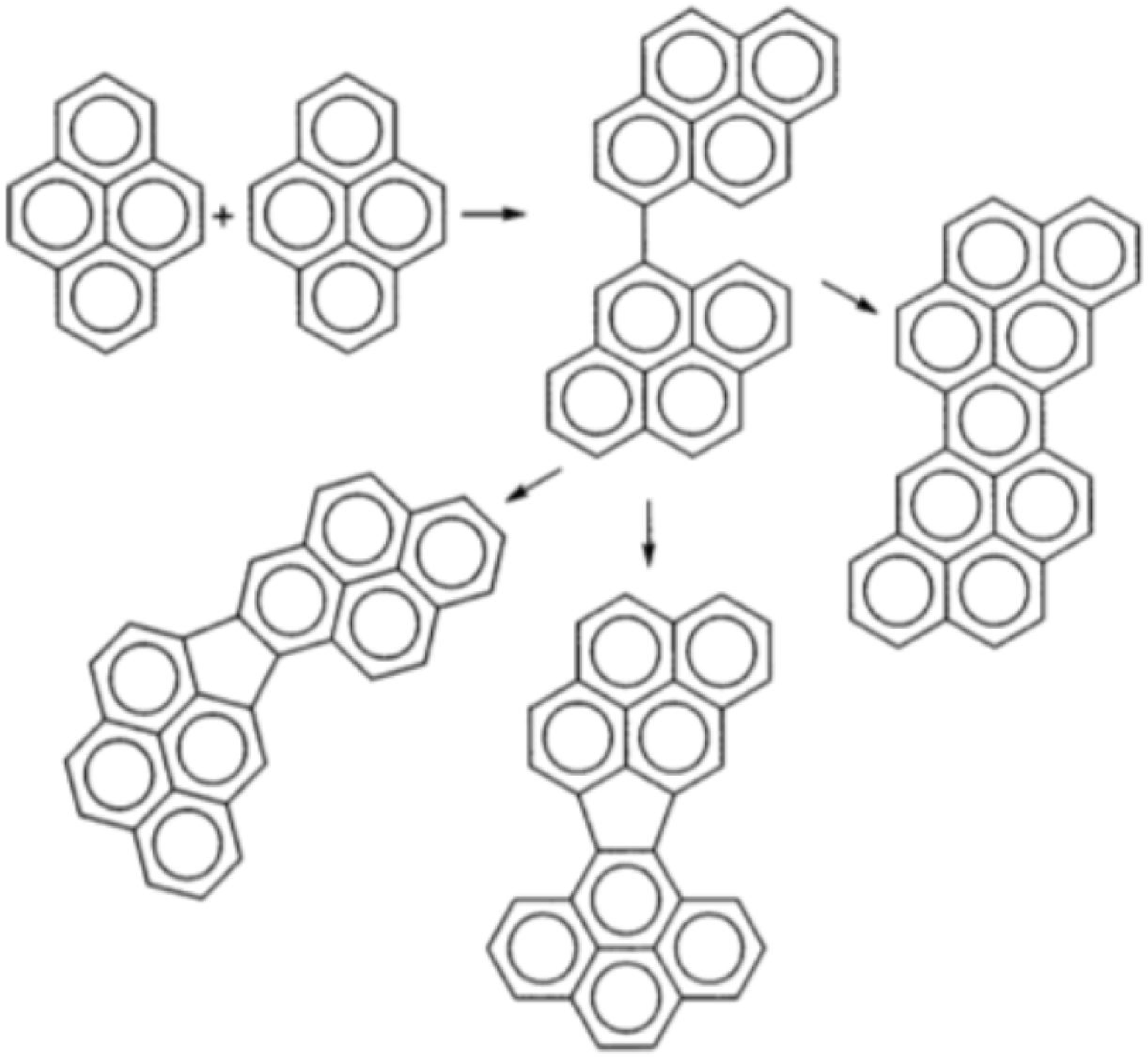} 
\caption{Left: Growth of PAHs according to the H abstraction-C$_2$H$_2$ addition (HACA) mechanism proposed by Frenklach et al. (1984). Right: Growth of aromatic structures via the dimerisation and coalescence of PAHs proposed by Mukhergee et al. (1994).} 
\end{figure}

Incomplete combustion produces soot particles where PAHs are observed to play a key role in the inception and growth process of AC grains (Richter \& Howard 2002). The formation of the single aromatic ring benzene (C$_6$H$_6$) represents the rate limiting step to soot synthesis as it is the natural passage from the aliphatic structure of the fuel hydrocarbons to the closed ring aromatic structures which are the backbones of combustion products like soot.
Three dominant routes to first ring closure have been identified in flames. The prevalent route involves the recombination of two propargyl radicals (C$_3$H$_3$) to form cyclic and linear  benzene and the benzene radical phenyl (C$_6$H$_5$) according to the following reactions
\begin{equation}
\label{R1}
 C_3H_3 + C_3H_3 \longrightarrow C_6H_6 
\end{equation}
and 
\begin{equation}
\label{R2}
 C_3H_3 + C_3H_3 \longrightarrow C_6H_5 + H 
\end{equation}
These routes were first studied and proposed by Miller \& Mellius (1992) as important sources of aromatic rings in flames. The propargyl recombination into one aromatic ring involves many reactive channels leading to the formation of several intermediates such as fulvene (C$_5$H$_4$CH$_2$), observed in flames (Miller \& Klippenstein 2003, Tang et al. 2006). In pyrolytic shock tube study of propargyl reactions, Scherer et al. (2000) show that the direct formation of benzene via Reaction \ref{R1} is prevalent over \ref{R2}. 

The other two chemical pathways to first aromatic ring closure in acetylenic flames involve the reaction of 1-buten-3-ynyl, 1-C$_4$H$_3$ with acetylene proposed by Frenklach et al. (1984), and the reaction of 1,3-butadienyl, 1-C$_4$H$_5$ with acetylene proposed by Cole et al. (1984). They lead to the following final products via direct pathways involving chemically activated isomerizations (Westmorland et al. 1989)
\begin{equation}
\label{R3}
 1-C_4H_3 + C_2H_2 \longrightarrow C_6H_5 
\end{equation}
and 
\begin{equation}
\label{R4}
 1-C_4H_5 + C_2H_2 \longrightarrow C_6H_6 + H. 
\end{equation}
 
In the context of PAH formation in circumstellar outflows, the above pathways will occur if the reactants are present in the gas. The formation of C$_3$H$_3$ results from the reaction of C$_2$H$_2$ and methylene, CH$_2$, whereas vynyl-acetylene, C$_4$H$_4$ will form C$_4$H$_3$ and C$_4$H$_5$ from its reaction with atomic hydrogen. 

Once benzene is made available in the gas, larger PAHs may grow according to three different scenarios. The first scenario was proposed by Frenklach et al. (1984) and involves the growth of large aromatic structures through sequential hydrogen abstraction and acetylene addition (referred as the HACA mechanism). The formation of radical sites on the aromatic structure enables C$_2$H$_2$ to add and finally form a new aromatic ring as depicted in Figure 1. The second route was proposed by Mukherjee et al. (1994) in their study of the pyrolysis of pyrene (C$_{14}$H$_{10}$). In the absence of PAH species arising from the HACA growth pathway, they derive that the dominant weight growth channel was the direct polymerisation of PAHs through the initial formation of PAH dimers, and the transformation of their van der Waals bonds into aliphatic bonds, as illustrated in Figure 1. Finally, a third growth pathway was proposed by Krestinin et al. (2000) in their study of acetylene pyrolysis and involves the polymerization of polyynes on a surface radical site of any small grain seed as illustrated in Figure 2. This hypothesis is supported by the observation of small polyynes (C$_4$H$_2$, C$_6$H$_2$, C$_8$H$_2$) along with PAHs in the sooting zone of flames (Cole et al. 1984). 

One should bear in  mind that there still exists controversy on the prevalent pathways to AC grain growth (for more detail, see J{\"a}ger et al., this volume). Despite the fact that the HACA mechanism is largely used to describe the growth of aromatic rings, it does not provide a satisfactory explanation for the early formation in sooting flames of small carbonaceous nanoparticles made of a few aromatic rings linked by aliphatic bonds (Minutolo et al. 1998). 

\begin{figure} 
\includegraphics[width=11cm]{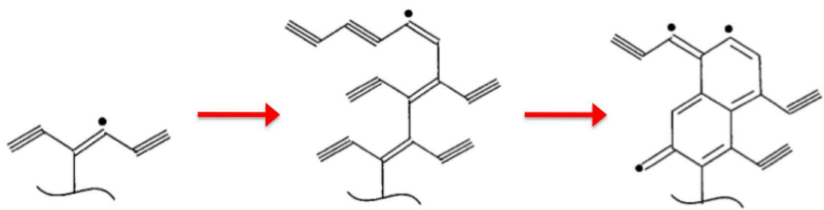}  
\caption{Polymerization of polyynes on a surface radical site and its transformation into aromatic structures leading to breeding of radical sites and rapid growth (from Krestinin 2000).} 
\end{figure}
\section{PAH yields from existing studies}

Only a few studies have been conducted to date on the formation of PAHs in the outflows of carbon AGB stars. Frenklach \& Feigelson (1989) first applied the HACA mechanism to the steady wind of carbon stars to derive dust masses and PAH yields. They highlighted the existence of a temperature window (900 K -1100 K) for which PAH growth occurred. However, their treatment of the stellar wind was not taking into account the action of shocks active above the stellar photosphere. Therefore, the gas density dropped by orders of magnitude over short time scales impeding the efficient synthesis of PAHs and dust grains. Cherchneff et al. (1992) studied the formation of PAHs using the HACA mechanism for PAH growth but including the newly proposed Reactions 5.1 and 5.2 for first ring closure. They include a treatment of the effects of shocks on the gas above the photosphere but also used a Eulerian treatment for the steady wind. The resulting PAH yields were small for all the various wind models and the shocked layers close to the stars experiencing quasi-ballistic trajectories were proposed as possible loci for a more efficient synthesis of PAHs. Cadwell et al. (1994) studied the induced nucleation  and growth of carbon dust on pre-existing seed nuclei. Finally Cherchneff \& Cau (1999) and Cau (2002) investigated the formation of PAHs and their dimers in the shocked inner wind of IRC+10216, using a Lagrangian formalism that follows the post-shock gas trajectories induced by the periodic shocks and the stellar gravitational field, following a study from Bowen (1988). The formation of PAHs and their dimers occurred at higher temperatures than those of Frenklach \& Feigelson, typically for T $\leq$ 1700 K, in agreement with a recent laboratory study of soot formation in low-temperature laser-induced pyrolysis of hydrocarbons by J{\"a}ger et al. (2009) . The deduced PAHs yields were sufficient to account for the mass of PAHs included in the nanocrystalline cores of graphite spherules. However, they could not account for the total mass of AC dust formed in the wind, indicating that further dust growth mechanisms must occur such as the deposition of acetylene on the surface of dust seeds. 

In view of these existing theoretical studies and new observational results, the ubiquity of PAHs in the inner wind of carbon AGB stars is almost certain. However, a complete model that includes the various pathways of PAH formation and growth to AC dust as described in Section 5, and applied to the shocked inner layers of carbon stars is still lacking.   

\section{New results on benzene formation in IRC+10216}
\begin{figure} 
\includegraphics[width=11.5cm]{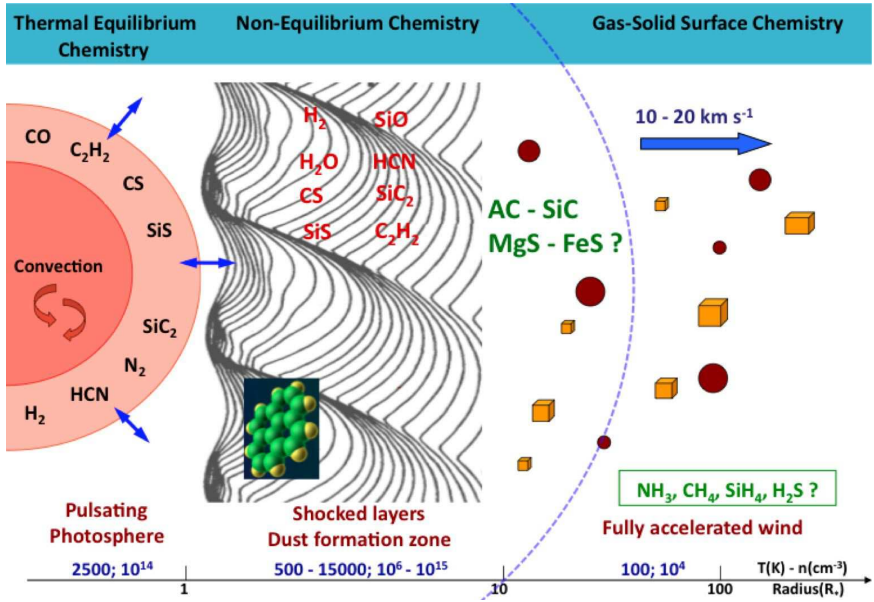}  
\caption{A schematic representation of the inner envelope of IRC+10216. The gas parameters span the values indicated as a function of radius. The quasi-balastic trajectories of the gas layers are shown according to Bowen (1988). Dust grains are represented by small squares and balls. The mentioned molecules form according to TE and non-TE chemistry depending on their locus. PAHs form in the inner wind between 1 \rstar\ and $\sim$ 5 \rstar.} 
\end{figure}

The carbon star IRC+10216 has been extensively studied due to its proximity and its stage of evolution that guarantees an extremely rich chemistry in its wind (Olofsson 2006). The dust forms close to the star in gas layers that experience the periodic passage of shocks induced by the stellar pulsation. The energy spectral distribution of IRC+10216 is well reproduced by a dust composition mainly dominated by amorphous carbon with a small contribution from silicon carbide and magnesium sulphide (Keady et al. 1988, Ivezi{\'c} \& Elitzur 1996, Hony et al. 2002). Acetylene has been detected in the mid-infrared close to the star with high abundances (Keady \& Ridgway 1993, Fonfr{\'i}a et al. 2008). A schematic representation of the inner wind of IRC+10216 is given in Figure 3 with the gas temperature and density as a function of radius, the various chemical states of the gas and the molecules observed. Briefly, molecules are formed at chemical equilibrium in the photosphere owing to the high gas temperatures and densities. They are then reprocessed by the periodic passage of shocks induced by the stellar pulsation. In the hot postshock gas, molecules are destroyed by collisions and reform during the adiabatic cooling of the gas as shown in Figure 3. Prevalent species (e.g. CO, HCN, CS, C$_2$H$_2$) reform with abundances close to their TE values whereas the collisional breaking of CO after the shock front induces a new chemistry responsible for the formation of 'exotic' molecules of lower abundances and not expected to form in C-rich environments. As an example, the radical hydroxyl, OH, quickly forms and triggers the synthesis of O-bearing species like SiO in the inner wind (Willacy \& Cherchneff 1998). Thus, an active non-equilibrium chemistry is responsible for the reprocessing of molecules in the dust formation zone of AGB stars (Cherchneff 2006). 
\begin{figure} 
\includegraphics[width=10.5cm]{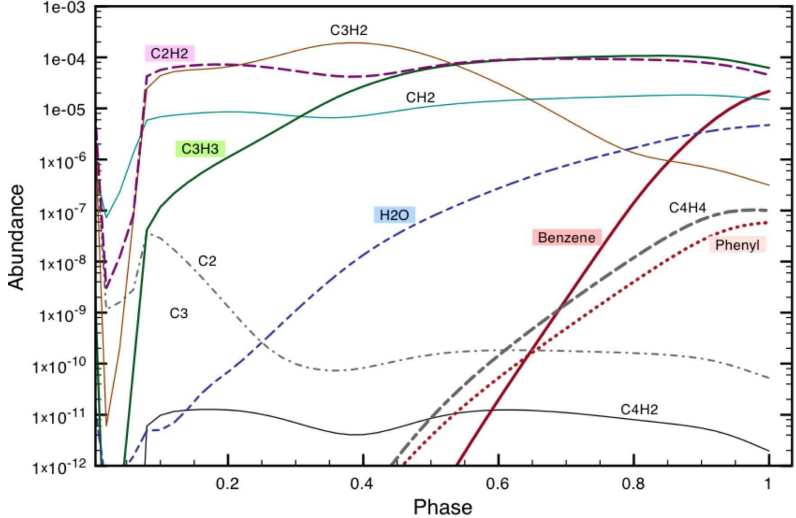}  
\caption{Molecular abundances with respect total gas as a function of pulsation period phase at 1.5 \rstar. The shock strength at this radius is 17.9 km s$^{-1}$.} 
\end{figure}

PAHs do not escape this shock molecular reprocessing and form once the adiabatic cooling of the shocked gas layers produces the temperature range conducive to the closure of the first aromatic ring (Cherchneff \& Cau 1998, Cau 2002). We model the inner wind of IRC+10216 using the Lagrangian formalism of Willacy \& Cherchneff (1998) but considering gas densities lower by one order of magnitude than their values for those were slightly overestimated (Ag{\'u}ndez \& Cernicharo 2006). An updated chemistry is used that includes the formation of the major O/Si/S/C-bearing molecules as well as the routes to the closure of the first aromatic ring described in Section 5. The results for the adiabatically-cooled post-shock gas layer located at 1.5 \rstar\ are shown in Figure 4, where molecular abundances with respect to total gas are presented versus phase of the pulsation period. As previously mentioned, benzene, C$_6$H$_6$ forms once the gas has cooled to temperatures $<$1700 K, while phenyl, C$_6$H$_5$ is consumed in benzene through its reaction with H$_2$. The formation of  C$_6$H$_5$ is triggered by the recombination of propargyl radicals, C$_3$H$_3$, according to Reaction \ref{R2}. The low abundances of C$_4$H$_2$ hint at the fact that PAH growth beyond benzene cannot proceed through polyyne polymerization but may rather involve the HACA mechanism and/or the direct coalescence of PAHs. As expected, the abundances of small carbon chains C$_2$ and C$_3$ peak at very early phases where the high temperatures precludes the formation of aromatics. However, their abundances are always low for the overabundance of hydrogen favours the formation of hydrocarbons over bare carbon chains. 

A secondary important result is the formation of water, H$_2$O, in the inner wind of IRC+10216. Our results show that H$_2$O already forms at radius 1.5 \rstar. OH is quickly synthesised after the shock front  at early phases by the reaction of atomic oxygen stemming from the partial dissociation of CO with the newly recombined H$_2$. H$_2$O is then synthesised from the reaction of OH with H$_2$. Water has recently been detected with Herschel/HIFI in the carbon star V Cygni (Neufeld et al. 2010), in the S star Chi Cygni (Justtanont et al. 2010), and finally in IRC+10216 (Decin et al. 2010). Those detections imply an inner origin for H$_2$O. Several mechanisms have been proposed since the first detection of H$_2$O in IRC+10216 (Melnick et al. 2001), and include the vaporisation of icy comets orbiting carbon stars (Melnick et al. 2001), the synthesis of water on dust grain surfaces in the intermediate envelope (Willacy 2004), and the photodissociation of CO by interstellar UV photons penetrating clumps (Decin et al. 2010). Our preliminary results point to a more genuine mechanism for water formation which results from the non-equilibrium chemistry induced by shocks in the inner wind of IRC+10216. The H$_2$O molecule may be added to the group of species shown by Cherchneff (2006) to form efficiently in the inner wind of AGB stars whatever their C/O ratio and stage of evolution on the AGB. The presence of water in the dust formation zone indicates that PAH formation and growth processes may be close to those found in acetylenic flames where oxidation by atomic O and O$_2$ occurs. While the amounts of atomic O and dioxygen remain low at all phases in the gas layers under study, OH may attack propargyl radicals to form the formyl radical HCO. Therefore, the benzene abundances and yields shown in Figure 3 may be overestimated due to the fact that the oxidation of hydrocarbons has not yet been included in the chemistry. Nevertheless, our preliminary results indicate a rapid conversion of hydrocarbons into monoaromatic rings. 

\section{Other environments conducive to PAH synthesis}

Apart from the winds of carbon AGB stars, there may exist other circumstellar media where PAHs form. One such environment is the wind of one hydrogen-rich R Corborealis star, V854 Centauri. R CrB stars are extremely rare, evolved, low-mass stars showing supergiant photospheric signatures. They have lost their entire hydrogen shell and experience strong fading on periods of the order of a few years. This sudden drop in luminosity is ascribed to the random ejection of AC dust clouds along the line of sight. V854 Cen is a peculiar R CrB as it has retained some hydrogen in its wind. Of particular interest is the detection of red emission (RE) bands at minimum whose wavelengths match those of a series of emission bands observed in the proto-planetary nebula, the Red Rectangle (Rao \& Lambert 1993). Furthermore, those frequencies almost match some of the Diffuse Interstellar Bands (DIBs) observed in the ISM.  The UIR bands were also detected with ISO in V854 Cen pointing to a PAH component in the stellar wind while the Swan and Phillip bands of C$_2$ were observed during the decline to light minimum (Rao \& Lambert 1993b, 2000, Lambert et al. 2001). The stellar temperature is high ($\sim$ 7500 K) indicating that despite the presence of hydrogen, the formation of bare carbon chains should proceed, as confirmed by the presence of C$_2$. As the temperature of the cloud drops with expansion and because of the surrounding hydrogen, polyyne molecules could form along with already formed carbon monocyclic and aromatic rings, leading to a growth mechanism for dust involving the polymerization of polyynes on radical sites (Krestinin et al. 2000). This growth mechanism should partly result in a left-over population of nanoparticles made of units of one or two aromatic rings linked by aliphatic bonds, as observed at high temperatures in sooting flames (Minutolo et al. 1998).  Hence the formation of PAHs in V854 Cen occurs along chemical pathways different from those involved in AGB winds (e.g., the HACA mechanism) and these aromatic nanoparticles could be serious contenders as carriers of the RE and the DIBs. 

\section{Conclusion}

The formation of PAHs from the gas phase and their role in the synthesis of carbon dust occurring in hydrogen-rich, low-temperature circumstellar environments like AGB stellar winds are well accepted. Small PAHs act as building blocks in the formation of carbonaceous nanoparticles precursors to dust seeds whereas large ones are direct intermediates to the synthesis of AC grains. A population of small PAHs will survive the AC dust condensation zone and is probably responsible for the UIR bands observed in certain AGB winds. However, most PAHs formed in the inner gas layers will be quickly included into AC dust grains and in this context, carbon AGB stars are not important PAH providers to the ISM. Despite the fact that circumstellar PAHs are less studied than their interstellar ubiquitous counterparts stemming from AC dust processing in the ISM, they are as important, being at the origin of the carbon dust formation process in evolved stars.    

%%-----------------------------
%%      your bibliography
%%-----------------------------

\end{document}